\documentclass[12pt]{article}
\usepackage{amsmath} 
\usepackage{longtable}
\usepackage{graphicx,epsfig}    
\usepackage{color}
\usepackage{cite} 
\usepackage{amssymb}
\usepackage[utf8]{inputenc}
\usepackage{appendix}
\usepackage{array}
\usepackage{mathtools}
\usepackage{xspace}
\usepackage{setspace}
\definecolor{darkblue}{RGB}{0, 56, 128}
\usepackage[hyperfootnotes=true,bookmarks=false,
hypertexnames=true,pagebackref=false,linktocpage=false,hidelinks]{hyperref}
\hypersetup{
   colorlinks=true,%
   linkcolor=darkblue,%
   citecolor=darkblue,%
   filecolor=green,%
   menucolor=cyan,%
   urlcolor=darkblue}

\usepackage{footnotebackref}

\newcommand{\as}{\left(\frac{\alpha_s}{ \pi}\right)}
\newcommand{\aq}{\left(\frac{\alpha}{ \pi}\right)}

\newcommand{\asa}{\alpha_s\alpha}
\newcommand{\qcdqed}{QCD$\otimes$QED\xspace}

\usepackage{array}
\newcolumntype{L}[1]{>{\raggedright\let\newline\\\arraybackslash\hspace{0pt}}m{#1}}
\newcolumntype{C}[1]{>{\centering\let\newline\\\arraybackslash\hspace{0pt}}m{#1}}
\newcolumntype{R}[1]{>{\raggedleft\let\newline\\\arraybackslash\hspace{0pt}}m{#1}}
\usepackage{siunitx}

\begin{document}
\begin{titlepage}

\begin{flushright}
ICAS 049/20 \\
MPP-2020-38
\end{flushright}

\renewcommand{\thefootnote}{\fnsymbol{footnote}}
\thispagestyle{empty}
\noindent

\vspace{0.5cm}

\begin{center}
{\bf \Large 
 Mixed \qcdqed corrections to exclusive Drell Yan \\[0.2cm]
 production using the $q_T$-subtraction method
 \\}
  \vspace{1.25cm}
{\large
Leandro Cieri$^{(a)}$\footnote{leandro.cieri@fi.infn.it},
Daniel de Florian$^{(b)}$\footnote{deflo@unsam.edu.ar},
Manuel Der$^{(b)}$\footnote{mder@unsam.edu.ar} and 
Javier Mazzitelli$^{(c)}$\footnote{jmazzi@mpp.mpg.de} \\
}
 \vspace{1.25cm}
 {$^{(a)}$ INFN, Sezione di Firenze, I-50019 Sesto Fiorentino, Florence, Italy\\[0.3cm]
$^{(b)}$International Center for Advanced Studies (ICAS), ICIFI and ECyT-UNSAM,\\ 
 25 de Mayo y Francia, (1650) Buenos Aires, Argentina \\[0.3cm]
$^{(c)}$Max-Planck-Institut f\"ur Physik, F\"ohringer Ring 6, 80805 M\"unchen, Germany\\
 }
  \vspace{1.5cm}
  \large {\bf Abstract}
  \vspace{-0.2cm}
\end{center}

\begin{quote}
\pretolerance 10000

In this work we extend the $q_T$-subtraction formalism, originally developed for QCD corrections, to the case of mixed \qcdqed corrections, and apply it to the fully exclusive calculation of the ${\cal{O}}(\asa)$ contribution to the production of an off-shell $Z$ boson in hadronic collisions. We present explicit results for the subtraction term and the hard factor, therefore providing all the ingredients needed for the application of the formalism up to ${\cal{O}}(\asa)$. To study the phenomenological impact we consider the decay of the off-shell $Z$ boson into a pair of neutrinos, and present kinematical distributions for the final-state leptons at LHC energies.

\end{quote}

\hfill

\end{titlepage}
\setcounter{footnote}{0}
\renewcommand{\thefootnote}{\arabic{footnote}}
%-----------------------------------------------------------%
%
\section{Introduction}
\label{sec:intro}
The Large Hadron Collider (LHC) at CERN is reaching, for many observables, an impressive accuracy. The precision of the LHC measurements will be further enhanced in the next runs and even more in the High-Luminosity upgrade of the LHC (HL-LHC). Theory must be ready for this appointment, producing equally accurate instruments in order to interpret the high-precision data. In this framework, QCD corrections play a crucial role. However, higher order perturbative QCD corrections for many observables and benchmark processes are of the same order of QED/EW or mixed QCD-EW theoretical predictions. Next-to-next-to-leading order (NNLO) QCD theoretical predictions (the standard of theoretical precision) for observables measured at the LHC are quite accurate but, in many cases, they are not sufficient to match the current accuracy at the experimental level and the new precision that will be reached in the following years. This scenario motivates a new theoretical effort to go beyond NNLO QCD corrections by including the first QED/EW corrections, mixed QCD-EW contributions and even the next QCD perturbative order: the next-to-next-to-next-to-leading order (${\rm N^3LO}$).

The Drell-Yan (DY) mechanism~\cite{Drell:1970wh} (a benchmark process at modern colliders) constitutes a clear example of the statements of a precision observable. This process offers the possibility of studying fundamental electroweak (EW) parameters in a clean and accurate way. It also provides strong tests for QCD predictions and stringent information to determine parton distribution functions (PDFs) with high accuracy.
The experimental precision for the DY mechanism at the LHC is at the percent level for the total cross section, and the differential distributions are also measured at an impressively high accuracy. The perturbative QCD corrections have been computed at next-to-leading order (NLO) in ref.~\cite{Altarelli:1979ub}, at NNLO for the inclusive cross section in refs.~\cite{Hamberg:1990np,vanNeerven:1991gh,Harlander:2002wh} and considering differential distributions in refs.~\cite{Catani:2009sm,Melnikov:2006di,Melnikov:2006kv, Gavin:2010az,Gavin:2012sy,Boughezal:2016wmq}. In addition, threshold expansions have been also presented at ${\rm N^3LL}$ accuracy in association with soft-virtual cross sections at ${\rm N^3LO}$ in refs.~\cite{Ahmed:2014cla,Catani:2014uta}. Very recently, the ${\rm N^3LO}$ QCD corrections have been obtained for the inclusive cross section for the production of a lepton pair via virtual photon exchange~\cite{Duhr:2020seh}.

However, computing several terms in the $\alpha_s$ expansion is not enough to reach the ultimate accuracy goal, since the EW coupling $\alpha$ satisfies $\alpha \sim \alpha_s^2$, and therefore NLO EW corrections, i.e. ${\cal O}(\alpha)$, are expected to be of the same order as the NNLO QCD contributions.

The calculation of NLO EW corrections for the DY process has been addressed in refs.~\cite{Dittmaier:2001ay,Baur:2004ig,CarloniCalame:2006zq} and \cite{Baur:2001ze,Baur:1997wa} for charged currents (CC) and neutral currents (NC), respectively.
In order to improve our understanding of these EW effects, the calculation of their first order QCD corrections, i.e. the ${\cal O}(\asa)$, becomes necessary.
These corrections represent the first term in the fixed order expansion that takes into account {\it mixed} effects from the strong and electroweak interactions.

Different approaches have been followed in the literature in order to approximately combine the QCD and EW corrections~\cite{Cao:2004yy,Balossini:2009sa,Adam:2008pc,Li:2012wna,Barze:2013fru}, by either assuming the full factorisation or the additive combination of the strong and electroweak contributions.
Partial exclusive results have been presented for the resonance region, by relying on the pole approximation \cite{Dittmaier:2014koa,Dittmaier:2014qza,Dittmaier:2015rxo}.  

A perturbative calculation of the Drell-Yan mechanism can be  characterised by the following subsets: on one hand, {\it purely factorisable} terms that arise due to
initial state ({\it production}, from the initial state partons) and final state ({\it decay}, from the final state leptons) emission
and, on the other hand, {\it non-factorisable} terms
originated by soft photon exchange between the production and the decay. 
The non-factorisable $\mathcal{O}(\asa)$ terms have been shown~\cite{Dittmaier:2014qza,Dittmaier:2014koa,Dittmaier:2015rxo} 
to have a negligible impact on the cross section, allowing to treat effectively  Drell-Yan  
in the (resonant) limit of the decoupling between the production and decay processes, at least for the achieved experimental accuracy.
Several steps towards the computation of the (inclusive) initial state QCD$\times$EW corrections have been recently carried out in an analytical way~\cite{Bonciani:2016wya,Bonciani:2016ypc,Bonciani:2019nuy}. The appearance of massive gauge bosons results in extra complications, so it seems natural to start by looking at the case of QED contributions instead. 

The first computation of the mixed  \qcdqed $\mathcal{O}(\asa)$ corrections
to the inclusive on-shell production of a $Z$ boson in hadronic collisions was achieved in ref.~\cite{deFlorian:2018wcj}, by profiting from the available NNLO pure QCD corrections via the so-called abelianisation techniques~\cite{deFlorian:2015ujt,deFlorian:2016gvk}. Those contributions were shown to be of the order of the NNLO QCD corrections for LHC energies, which makes them relevant to reach an accurate theoretical description. 
Moreover, it would be highly desirable to evaluate their effect at a fully exclusive level.

A crucial ingredient in the calculation of fully differential distributions are the so-called subtraction methods.
For the case of pure QCD corrections to the hadroproduction of colourless final states, the $q_T$-subtraction method~\cite{Catani:2007vq,Catani:2013tia} has been extensively used in order to obtain NNLO-accurate predictions.
In this work, we extend the $q_T$-subtraction formalism in order to apply it to the calculation of ${\cal O}(\asa)$ mixed corrections at a fully exclusive level. Our results are of value for transverse-momentum resummation at the corresponding logarithmic accuracy.

In particular, we will focus on the mixed \qcdqed corrections to the production of an off-shell $Z$ boson decaying into a neutrino-antineutrino system. We consider the simplest case of uncharged particles in the decay of the off-shell $Z$ boson as a way to directly address the relevance on initial state corrections to a number of exclusive observables.
Note that a recent work~\cite{Delto:2019ewv} also considers the production of a $Z$ boson, though in this case on-shell, in a fully exclusive way, based on the abelianised version of the nested soft-collinear subtraction formalism~\cite{Caola:2017dug}.

This paper is organised as follows:
In Section~\ref{sec:Method} we present the relevant formulae for the extension of the $q_T$-subtraction method to the \qcdqed case. In Section~\ref{sec:results} we present our numerical results and study the phenomenology of the corrections for different kinematical variables. Finally, in Section~\ref{sec:conc} we present our conclusions.

\section{Mixed order corrections with $q_T$-subtraction formalism}
\label{sec:Method}

We consider the inclusive hard-scattering reaction
\begin{equation}
h_1(p_1)+h_2(p_2) \rightarrow F(M,q_T)+X\, ,
\label{eq:class}
\end{equation}
where the collision of the two hadrons $h_1$ and $h_2$ with momenta $p_1$ and
$p_2$ produces the triggered generic final state $F$, without colour and electric charge, such as \textit{one} or \textit{more} neutral vector bosons ($\gamma^{*},Z,ZZ,\gamma\gamma$, $\ldots$), Higgs particles, and so forth. The observed final state $F$ is accompanied by the final-state radiation $X$, which in this case, consists of either quarks, antiquarks, gluons or photons.
The system $F$ is composed by $n$ final-state particles with momenta $q_1,q_2,\dots,q_n$, and has {\em total} invariant mass $M^2=(q_1+q_2+\dots +q_n)^2$, transverse momentum $q_T$ and rapidity $y$. We use $\sqrt{s}$ to denote the centre-of-mass energy of the colliding hadrons, which are treated in the massless approximation ($s=(p_1+p_2)^2=2p_1\cdot p_2$).

We start by considering the \qcdqed perturbative expansion of the (differential) cross section for the production of the final state $F$, by expanding in powers of the strong ($\alpha_s$) and electromagnetic ($\alpha$) couplings,
\begin{equation}
\label{eq:expansion}
d\sigma^F = \sum _{i,j} \as^i \aq^j d\sigma^{(i,j)}_F,
\end{equation}
where $d\sigma^{(i,0)}_F$ stands for the pure QCD corrections, and $d\sigma^{(0,j)}_F$  for the pure QED ones. The \textit{mixed} corrections are represented by $d\sigma^{(i,j)}_F$ with both $i,j \neq 0$, being the first mixed contribution $d\sigma^{(1,1)}_F$. 

Following a similar structure to the one valid in the pure QCD case~\cite{Catani:2007vq,Catani:2013tia}, the basic formula for the ${q_T}$-subtraction method in the case of mixed \qcdqed corrections can be expressed in the following way,
\begin{equation}
\label{eq:qt}
    d\sigma_F^{(1,1)} = {\cal{H}}_F^{(1,1)} \otimes d \sigma_{F}^{(0,0)} + \left[ d\sigma^{(1,1)}_{F+{\rm jet}}-d\sigma^{(1,1)}_{F\, CT} \right]\,,
\end{equation}
where $d\sigma^{(1,1)}_{F+{\rm jet}}$ corresponds to the $F+\rm{jet}$ production cross section at ${\cal O}(\asa)$.
It is important to note that in this context `$\rm{jet}$' stands for either quarks, antiquarks, gluons or photons in the final state and all of them need to be considered in the initial state as well.
The term inside the square bracket in eq.~(\ref{eq:qt}) is finite in the limit of vanishing transverse momentum of the $F$ state, but the individual terms $d\sigma^{(1,1)}_{F+{\rm jet}}$ and $d\sigma^{(1,1)}_{F\, CT}$ are separately divergent.
In order to evaluate $d\sigma^{(1,1)}_{F+{\rm jet}}$, we can make use of any NLO subtraction method (adapted, though, to the case of mixed \qcdqed corrections). 

The subtraction counter-term $d\sigma^{(1,1)}_{F\,CT}$ encodes the singular behaviour of the real scattering amplitudes in the small-$q_{T}$ region. The coefficient function ${\cal{H}}_F^{(1,1)}$ restores the correct normalisation to the total cross section and it has Born kinematics (\textit{e.g.} it is proportional to $\delta(q_T)$). Both coefficient functions can be obtained, through the abelianisation procedures \cite{deFlorian:2016gvk,deFlorian:2015ujt}, from eqs.~(63-70) in ref.~\cite{Bozzi:2005wk}. We have checked (as a self-consistency check) that the same coefficient functions can be obtained from first principles, \textit{i.e} redefining eq.~(6) in ref. \cite{Catani:2013tia} to take into account QED emissions and expanding it to a given fixed order.

We present in the following the explicit expression of all the required terms needed for the subtraction at ${\cal O}(\asa)$. These are constructed by convoluting the parton distributions with the corresponding partonic terms, which up to ${\cal O}(\asa)$ are given by 
\begin{align}
 d \sigma_{ ab \, CT}^{F}
 & =\as \, d\sigma^{F\,(1,0)}_{ab\, CT} + \aq \, d\sigma^{F\,(0,1)}_{ab\, CT} + \as \aq \, d\sigma^{F\,(1,1)}_{ab\, CT} \notag\\[1.2ex]
 &=\sum_{c} d\sigma_{c \overline{c}, F}^{(0,0)} \Bigg\{  
\left(\frac{\alpha_{s}}{\pi}\right)\widetilde{\Sigma}_{c \overline{c} \leftarrow a b}^{F(1,0)}\left(z,q_T /Q\right)
+\left(\frac{\alpha}{\pi}\right)\widetilde{\Sigma}_{c \overline{c} \leftarrow a b}^{F(0,1)}\left(z,q_T /Q\right)  \\[1.ex]&
+\left(\frac{\alpha_{s}}{\pi}\right)\left(\frac{\alpha}{\pi}\right)\widetilde{\Sigma}_{c \overline{c} \leftarrow a b}^{F(1,1)}\left(z,q_T /Q \right)
\Bigg\}\notag
\end{align}
and
\begin{align}
\mathcal{H}_{ ab}^{F} \otimes d \sigma_{L O}^{F}& =\left[1+\as \, \mathcal{H}_{ ab}^{F(1,0)} + \aq \, \mathcal{H}_{ ab}^{F(0,1)} + \as \aq \, \mathcal{H}_{ ab}^{F(1,1)}\right]\otimes d \sigma_{L O}^{F}\notag\\[1.2ex]
& =\sum_{c} d\sigma_{c \overline{c}, F}^{(0,0)}\Bigg\{ \delta_{c a} \delta_{\overline{c} b} \delta(1-z) 
+\left(\frac{\alpha_{s}}{\pi}\right)
\mathcal{H}_{c \overline{c} \leftarrow a b}^{F(1,0)}\left(z \right)  \\[1.ex]&
+\left(\frac{\alpha}{\pi}\right)
\mathcal{H}_{c \overline{c} \leftarrow a b}^{F(0,1)}\left(z \right) 
+\left(\frac{\alpha_{s}}{\pi}\right)\left(\frac{\alpha}{\pi}\right)\mathcal{H}_{c \overline{c} \leftarrow a b}^{F(1,1)}\left(z \right) \Bigg\}\,.\notag
\end{align}
In order to simplify the notation, we indicate by $z$ the dependence on both partonic momentum fractions $z_1$ and $z_2$. The explicit dependence on either $z_1$ and $z_2$ can be easily understood in terms on the dependence on the partonic label $a$ and $b$, respectively.  Also, it is implicit the dependence on the renormalisation ($\mu_R$), factorisation ($\mu_F$) and resummation ($Q$) scales. 

Note that, for the sake of generality, in the results contained in this section we keep the full dependence on the resummation scale~\cite{Bozzi:2005wk}. This dependence is needed in the context of transverse-momentum resummation. The fixed-order cross-section is independent of this scale, and it is convenient to set $Q=M$ to simplify the corresponding expressions.

The contributions to the counter-term $\widetilde{\Sigma}_{c \overline{c} \leftarrow a b}^{F(i,j)}$ can be organized in the following way
\begin{equation}
\widetilde{\Sigma}_{c \overline{c} \leftarrow a b}^{F(1,0)}(z,q_T /Q)=\Sigma_{c \overline{c} \leftarrow a b}^{F(1,0)[1 ; 2]}(z) \widetilde{I}_{2}\left(q_{T} /Q\right)+\Sigma_{c \overline{c} \leftarrow a b}^{F(1,0)[1 ; 1]}(z) \widetilde{I}_{1}\left(q_{T} /Q\right)\,,
\end{equation}
\begin{equation}
\widetilde{\Sigma}_{c \overline{c} \leftarrow a b}^{F(0,1)}(z, q_T /Q)=\Sigma_{c \overline{c} \leftarrow a b}^{F(0,1)[1 ; 2]}(z) \widetilde{I}_{2}\left(q_{T} /Q\right)+\Sigma_{c \overline{c} \leftarrow a b}^{F(0,1)[1 ; 1]}(z) \widetilde{I}_{1}\left(q_{T} /Q\right)\,,
\end{equation}
\begin{align}
\widetilde{\Sigma}_{c \overline{c} \leftarrow a b}^{F(1,1)}(z, q_T /Q)&=\Sigma_{c \overline{c} \leftarrow a b}^{F(1,1)[2 ; 4]}(z) \widetilde{I}_{4}\left(q_{T} /Q\right)+\Sigma_{c \overline{c} \leftarrow a b}^{F(1,1)[2 ; 3]}(z) \widetilde{I}_{3}\left(q_{T} /Q\right) \notag\\[1.2ex]
&\,+\Sigma_{c \overline{c} \leftarrow a b}^{F(1,1)[2 ; 2]}(z) \widetilde{I}_{2}\left(q_{T} /Q\right)+\Sigma_{c \overline{c} \leftarrow a b}^{F(1,1)[2 ; 1]}(z) \widetilde{I}_{1}\left(q_{T} /Q\right) \, ,
\end{align}
according to their power of logarithmic enhancement.
The dependence on the transverse momentum is given by the known integrals~\cite{Bozzi:2005wk}
\begin{equation}
\label{eq:Itilde}
\widetilde{I}_{n}\left(q_{T} /Q\right)=Q^{2} \int_{0}^{\infty} d b \frac{b}{2} J_{0}\left(b q_{T}\right) \ln ^{n}\left(\frac{Q^{2} b^{2}}{b_{0}^{2}}+1\right) \, ,
\end{equation}
where $b$ is the impact parameter, $J_0(x)$ is the 0th-order Bessel function and $b_0 = 2 e^{-\gamma_E}$, with $\gamma_E$ representing the Euler number. Notice that we are using the ``$+1$'' prescription (see the argument of the logarithm inside eq.~(\ref{eq:Itilde})), and therefore, the counter-terms vanish in the large-$q_T$ limit. More details about eq.~(\ref{eq:Itilde}) can be found in the Appendix A of ref.~\cite{Bozzi:2005wk}. 

The corresponding coefficients for the expansion of $\widetilde{\Sigma}_{c \overline{c} \leftarrow a b}^{F(i,j)}$ and $\mathcal{H}_{c \overline{c} \leftarrow a b}^{F(i,j)}$ are more easily presented by considering their $N$-moments (Mellin) with respect to the variable $z$. At NLO in QCD and QED they are given by 
\begin{equation}
\label{eq:Sigma10_12}
\Sigma_{c \overline{c} \leftarrow a b, N}^{F(1,0)[1 ; 2]}=-\frac{1}{2} A_{c}^{(1,0)} \delta_{c a} \delta_{\overline{c} b}\,,
\end{equation}

\begin{equation}
\label{eq:Sigma10_11}
\Sigma_{c \overline{c} \leftarrow a b, N}^{F(1,0)[1 ; 1]}=-\left[\delta_{c a} \delta_{\overline{c} b}\left(B_{c}^{(1,0)}+A_{c}^{(1,0)} \ell_{Q}\right)+\delta_{c a} \gamma_{\overline{c} b, N}^{(1,0)}+\delta_{\overline{c}b} \gamma_{c a, N}^{(1,0)}\right]\,,
\end{equation}

\begin{equation}
\label{eq:Sigma01_12}
\Sigma_{c \overline{c} \leftarrow a b, N}^{F(0,1)[1 ; 2]}=-\frac{1}{2} A_{c}^{(0,1)} \delta_{c a} \delta_{\overline{c} b}\,,
\end{equation}

\begin{equation}
\label{eq:Sigma01_11}
\Sigma_{c \overline{c} \leftarrow a b, N}^{F(0,1)[1 ; 1]}=-\left[\delta_{c a} \delta_{\overline{c} b}\left(B_{c}^{(0,1)}+A_{c}^{(0,1)} \ell_{Q}\right)+\delta_{c a} \gamma_{\overline{c} b, N}^{(0,1)}+\delta_{\overline{c}b} \gamma_{c a, N}^{(0,1)}\right]\,,
\end{equation}

\begin{equation}
\begin{aligned}
\label{eq:H_10}
 \mathcal{H}_{c \overline{c} \leftarrow a b, N}^{F(1,0)} &=\delta_{c a} \delta_{\overline{c} b}\left[H_{c}^{F(1,0)}-\left(B_{c}^{(1,0)}+\frac{1}{2} A_{c}^{(1,0)} \ell_{Q}\right) \ell_{Q}\right] \\[1.3ex] &+\delta_{c a} C_{\overline{c} b, N}^{(1,0)}+\delta_{\overline{c }b} C_{c a, N}^{(1,0)}+\left(\delta_{c a} \gamma_{\overline{c} b, N}^{(1,0)}+\delta_{\overline{c}b} \gamma_{c a, N}^{(1,0)}\right)\left(\ell_{F}-\ell_{Q}\right) \,,
\end{aligned}
\end{equation}

\begin{equation}
\begin{aligned}
\label{eq:H_01}
 \mathcal{H}_{c \overline{c} \leftarrow a b, N}^{F(0,1)} &=\delta_{c a} \delta_{\overline{c} b}\left[H_{c}^{F(0,1)}-\left(B_{c}^{(0,1)}+\frac{1}{2} A_{c}^{(0,1)} \ell_{Q}\right) \ell_{Q}\right] \\[1.3ex] &+\delta_{c a} C_{\overline{c} b, N}^{(0,1)}+\delta_{\overline{c} b} C_{c a, N}^{(0,1)}+\left(\delta_{c a} \gamma_{\overline{c} b, N}^{(0,1)}+\delta_{\overline{c }b} \gamma_{c a, N}^{(0,1)}\right)\left(\ell_{F}-\ell_{Q}\right) \,,
\end{aligned}
\end{equation}
while for the mixed \qcdqed corrections at ${\cal O}(\asa)$ they are given by
\begin{equation}
\label{eq:Sig11_24}
\Sigma_{c \overline{c} \leftarrow a b, N}^{F(1,1)[2 ;4]}=\frac{1}{4} A_{c}^{(1,0)} A_{c}^{(0,1)}\delta_{c a} \delta_{\overline{c} b}\,,
\end{equation}

\begin{equation}
\label{eq:Sig11_23}
\Sigma_{c \overline{c}\leftarrow a b, N}^{F(1,1)[2;3]}=-A_{c}^{(0,1)} \frac{1}{2} \Sigma_{c \overline{c}-a b, N}^{F(1,0)[1 ; 1]}-A_{c}^{(1,0)} \frac{1}{2} \Sigma_{c \overline{c}-a b, N}^{F(0,1)[1 ; 1]}\,,
\end{equation}

\begin{align}
\label{eq:Sig11_22}
\Sigma_{c \overline{c}\leftarrow a b, N}^{F(1,1)[2 ; 2]} =&
-\frac{1}{2} A_{c}^{(1,0)}\mathcal{H}_{c \overline{c}\leftarrow a b, N}^{F(0,1)} 
-\frac{1}{2} \sum_{a_{1}, b_{1}} \Sigma_{c \overline{c}\leftarrow a_{1} b_{1}, N}^{F(1,0)[1 ; 1]}\left[\delta_{a_{1} a} \gamma_{b_{1} b, N}^{(0,1)}+\delta_{b_{1} b} \gamma_{a_{1} a, N}^{(0,1)}\right]\notag\\
&-\frac{1}{2} A_{c}^{(0,1)}\mathcal{H}_{c \overline{c}\leftarrow a b, N}^{F(1,0)} 
-\frac{1}{2} \sum_{a_{1}, b_{1}} \Sigma_{c \overline{c}\leftarrow a_{1} b_{1}, N}^{F(0,1)[1 ; 1]}\left[\delta_{a_{1} a} \gamma_{b_{1} b, N}^{(1,0)}+\delta_{b_{1} b} \gamma_{a_{1} a, N}^{(1,0)}\right] \\
 &-\frac{1}{2}\left[A_{c}^{(1,1)} \delta_{c a} \delta_{\overline{c} b}+\left(B_{c}^{(1,0)}
 +A_{c}^{(1,0)} \ell_{Q}\right) \Sigma_{c \overline{c}\leftarrow a b, N}^{F(0,1)[1 ; 1]}
+\left(B_{c}^{(0,1)}+A_{c}^{(0,1)} \ell_{Q}\right) \Sigma_{c \overline{c}\leftarrow a b, N}^{F(1,0)[1 ; 1]}
\right] \,,\notag
\end{align}

\begin{equation}
\label{eq:Sig11_21}
\begin{aligned}
 \Sigma_{c \overline{c}\leftarrow a b, N}^{F(1,1)[2 ; 1]} =&
  -\sum_{a_{1}, b_{1}} \mathcal{H}_{c \overline{c} \leftarrow a_{1} b_{1}, N}^{F(1,0)}\left[\delta_{a_{1} a} \delta_{b_{1} b}\left(B_{c}^{(0,1)}+A_{c}^{(0,1)} \ell_{Q}\right)+\delta_{a_{1} a} \gamma_{b_{1} b, N}^{(0,1)}+\delta_{b_{1} b} \gamma_{a_{1} a, N}^{(0,1)}\right] \\
   &-\sum_{a_{1}, b_{1}} \mathcal{H}_{c \overline{c}\leftarrow a_{1} b_{1}, N}^{F(0,1)}\left[\delta_{a_{1} a} \delta_{b_{1} b}\left(B_{c}^{(1,0)}+A_{c}^{(1,0)} \ell_{Q}\right)+\delta_{a_{1} a} \gamma_{b_{1} b, N}^{(1,0)}+\delta_{b_{1} b} \gamma_{a_{1} a, N}^{(1,0)}\right] \\
&-\left[\delta_{c a} \delta_{\overline{c} b}\left(B_{c}^{(1,1)}+A_{c}^{(1,1)} \ell_{Q}\right)
  +\delta_{c a} \gamma_{\overline{c} b, N}^{(1,1)}+\delta_{\overline{c} b} \gamma_{ca, N}^{(1,1)}\right]\,,
  \end{aligned}
\end{equation}

\begin{spreadlines}{2ex}
\begin{align}
\label{eq:H11}
\mathcal{H}_{c \overline{c}\leftarrow a b, N}^{F(1,1)}&=\delta_{c a} \delta_{\overline{c} b} H_{c}^{F(1,1)}+\delta_{c a} C_{\overline{c} b, N}^{(1,1)}+\delta_{\overline{c} b} C_{c a, N}^{(1,1)}
+C_{c a, N}^{(1,0)} C_{\overline{c} b, N}^{(0,1)}+C_{c a, N}^{(0,1)} C_{\overline{c} b, N}^{(1,0)}
 \notag\\
 &+H_{c}^{F(1,0)}\left(\delta_{c a} C_{\overline{c} b, N}^{(0,1)}+\delta_{\overline{c} b} C_{c a, N}^{(0,1)}\right)
 +H_{c}^{F(0,1)}\left(\delta_{c a} C_{\overline{c} b, N}^{(1,0)}+\delta_{\overline{c} b} C_{c a, N}^{(1,0)}\right) \notag\\
& +\frac{1}{2}A_{c}^{(1,1)} \delta_{c a} \delta_{\overline{c} b} \ell_{Q}^{2} +\left(\delta_{c a} \gamma_{\overline{c} b, N}^{(1,1)}+\delta_{\overline{c} b} \gamma_{c a, N}^{(1,1)}\right) \ell_{F} \notag\\
&-\left[\delta_{c a} \delta_{\overline{c}b}\left(B_{c}^{(1,1)}+A_{c}^{(1,1)} \ell_{Q}\right)+\delta_{c a} \gamma_{\overline{c} b, N}^{(1,1)}+\delta_{\overline{c}b } \gamma_{c a, N}^{(1,1)}\right] \ell_{Q} \notag\\
&+\frac{1}{2} \sum_{a_{1}, b_{1}}\left[\mathcal{H}_{c \overline{c} \leftarrow a_{1} b_{1}, N}^{F(1,0)}
+\delta_{c a_{1}} \delta_{\overline{c} b_{1}} H_{c}^{F(1,0)}+\delta_{c a_{1}} C_{\overline{c} b_{1}, N}^{(1,0)}+\delta_{\overline{c} b_{1}} C_{c a_{1}, N}^{(1,0)}\right]
\\
 &\;\;\;\;\; \times\left[\left(\delta_{a_{1} a} \gamma_{b_{1} b, N}^{(0,1)}+\delta_{b_{1} b} \gamma_{a_{1} a, N}^{(0,1)}\right)\left(\ell_{F}-\ell_{Q}\right)-\delta_{a_{1} a} \delta_{b_{1} b}\left(\left(B_{c}^{(0,1)}+\frac{1}{2} A_{c}^{(0,1)} \ell_{Q}\right) \ell_{Q}\right)\right] \notag\\
 &+\frac{1}{2} \sum_{a_{1}, b_{1}}\left[\mathcal{H}_{c \overline{c} \leftarrow a_{1} b_{1}, N}^{F(0,1)}
+\delta_{c a_{1}} \delta_{\overline{c} b_{1}} H_{c}^{F(0,1)}+\delta_{c a_{1}} C_{\overline{c} b_{1}, N}^{(0,1)}+\delta_{\overline{c} b_{1}} C_{c a_{1}, N}^{(0,1)}\right]
\notag\\
 &\;\;\;\;\; \times\left[\left(\delta_{a_{1} a} \gamma_{b_{1} b, N}^{(1,0)}+\delta_{b_{1} b} \gamma_{a_{1} a, N}^{(1,0)}\right)\left(\ell_{F}-\ell_{Q}\right)-\delta_{a_{1} a} \delta_{b_{1} b}\left(\left(B_{c}^{(1,0)}+\frac{1}{2} A_{c}^{(1,0)} \ell_{Q}\right) \ell_{Q}\right)\right] \,.\notag
 \end{align}
 \end{spreadlines}

In the expressions above we have defined $\ell_{Q}=\ln M^2/Q^2$ and $\ell_{F}=\ln M^2/\mu_F^2$, while $ \gamma_{ab, N}^{(i,j)}$ represent the corresponding (moments of the) splitting functions.
The coefficients $A_c^{(i,j)}$ and $B_c^{(i,j)}$ arise from the expansion of the Sudakov form factor,
\begin{equation}
S_{c}(M, b)=\exp \left\{-\int_{b_{0}^{2} / b^{2}}^{M^{2}} \frac{d q^{2}}{q^{2}}\left[A_{c}\left(\alpha_{s},\alpha\right) \ln \frac{M^{2}}{q^{2}}+B_{c}\left(\alpha_{s},\alpha\right)\right]\right\}\,,
\end{equation}
with
\begin{equation}
\begin{aligned} 
A_{c}\left(\alpha_{s},\alpha\right) &=\left(\frac{\alpha_{s}}{\pi}\right) A_{c}^{(1,0)} +
\left(\frac{\alpha}{\pi}\right) A_{c}^{(0,1)}+
\left(\frac{\alpha_{s}}{\pi}\right)\left(\frac{\alpha}{\pi}\right) A_{c}^{(1,1)} + \dots \,,  \\[2ex] 
B_{c}\left(\alpha_{s},\alpha\right) &=\left(\frac{\alpha_{s}}{\pi}\right) B_{c}^{(1,0)} +
\left(\frac{\alpha}{\pi}\right) B_{c}^{(0,1)}+
\left(\frac{\alpha_{s}}{\pi}\right)\left(\frac{\alpha}{\pi}\right) B_{c}^{(1,1)} + \dots \,,
\end{aligned}
\end{equation}
and their explicit expression for quark-initiated case is given by
\begin{eqnarray}
&&A_{q}^{(1,0)}=C_F\, ,  \,\,\,\,\,\,\,\,\, A_{q}^{(0,1)}=e_q^2 \,, \nonumber \\[1ex]
&&B_{q}^{(1,0)}=-\frac{3}{2}C_F\, ,  \,\,\,\,\,\,\,\,\, B_{q}^{(0,1)}=-\frac{3}{2}e_q^2 \,, \\[1ex]
&&A_{q}^{(1,1)}=0 \, ,  \,\,\,\,\,\,\,\,\, B_{q}^{(1,1)}=\frac{C_F e_q^2}{8} (-3+24 \zeta_2-48\zeta_3) \,. \nonumber
\end{eqnarray}
Notice that we consider the electromagnetic coupling $\alpha$ as constant, in the sense that it is not running with any of the scales related to the process. For that reason the QED beta-function does not appear in the coefficients of eqs.~(\ref{eq:Sigma10_12}--\ref{eq:H11}). 
Eqs.~(\ref{eq:Sigma10_12}--\ref{eq:H_01}) were derived for first time in ref.~\cite{Cieri:2018sfk}, where the transverse-momentum resummation for $Z$ boson production combining QED and QCD was computed at NLO. 
It is worth noticing that for transverse-momentum resummation some novel mixed effects appear affecting the distribution already at leading logarithmic (LL) accuracy (see eqs.~(7) and (11) of ref.~\cite{Cieri:2018sfk}).
Nevertheless, that contribution can only show up after performing the fixed order expansion up to $\mathcal{O}(\asa)$ (see eq.~(3) in ref.~\cite{Cieri:2018sfk}) if the electromagnetic coupling $\alpha$ is considered to be running, which is not the case in our current study.

Finally we present the collinear functions, again for $c=q$, and the hard-virtual coefficients, the latter specifically for the DY case as they are a process-dependent quantity. The separation between $C$ and $H$ coefficients is scheme dependent. Those presented here are obtained in the so-called {\it hard} scheme~\cite{Catani:2013tia}.
Up to NLO in QCD and QED, the hard-virtual coefficients take the form
\begin{eqnarray}
H_{q}^{DY(1,0)}= C_F \left(\frac{\pi^{2}}{2}-4\right) = \frac{C_F}{e_q^2} H_{q}^{DY(0,1)}\,,
\end{eqnarray}
 and the collinear functions are given by
\begin{spreadlines}{1.5ex}
\begin{align}
C_{q q}^{(1,0)}(z)&=\frac{C_F}{2} (1-z)=\frac{C_F}{e_q^2} C_{q q}^{(0,1)}(z)\,,\nonumber \\
C_{q g}^{(1,0)}(z)&=\frac{1}{2} z(1-z)=C_{q \gamma}^{(0,1)}(z) \frac{T_R}{e_q^2 N_C}\,, \\
C_{g q}^{(1,0)}(z)&=\frac{C_F}{2} z=\frac{C_F}{e_q^2} C_{\gamma q}^{(0,1)}(z)\,. \nonumber
\end{align}
\end{spreadlines}
The hard-virtual coefficient needed for the first order in the mixed \qcdqed expansion takes the following form,
\begin{eqnarray}
H_{q}^{DY(1,1)}= \frac{C_F e_q^2}{2} \left(-15 \zeta_{3}+\frac{511}{16}-\frac{67 \pi^{2}}{12}+\frac{17 \pi^{4}}{45}\right)\,,
\end{eqnarray}
while the needed collinear functions are given by the following expressions:
\begin{spreadlines}{1.5ex}
\begin{align}
C_{q q'}^{(1,1)}(z) &=\delta_{q q'}\, e_q^2 C_{F} \Bigg\{
\frac{1+z^{2}}{1-z}\Bigg(\frac{\operatorname{Li}_{3}(1-z)}{2}+\frac{1}{2} \operatorname{Li}_{2}(z) \log (1-z)+\frac{3 \operatorname{Li}_{2}(z) \log (z)}{2} \nonumber \\ 
&-\frac{5 \operatorname{Li}_{3}(z)}{2} +\frac{3}{4} \log (z) \log ^{2}(1-z)+\frac{1}{4} \log ^{2}(z) \log (1-z)-\frac{1}{12} \pi^{2} \log (1-z)+\frac{5 \zeta_{3}}{2}\Bigg) \nonumber  \\ 
&+(1-z)\left(-\operatorname{Li}_{2}(z)-\frac{3}{2} \log (1-z) \log (z)+\frac{2 \pi^{2}}{3}-\frac{29}{4}\right)+\frac{1}{24}(1+z) \log ^{3}(z) \\ 
&+\frac{1}{1-z}\left(\frac{1}{8}\left(-2 z^{2}+2 z+3\right) \log ^{2}(z)+\frac{1}{4}\left(17 z^{2}-13 z+4\right) \log (z)\right) \nonumber \\
&-\frac{z}{4} \log (1-z) -\frac{1}{4} \left[ (2 \pi^2-18)(1-z)-(1+z)\log z  \right]
\Bigg\} \,,\nonumber
\end{align}
\end{spreadlines}
\begin{spreadlines}{1.5ex}
\begin{align} 
C_{q \overline{q}'}^{(1,1)}(z)&=\delta_{q q'}\, 2 C_{F}  e_q^2\left\{\frac{1+z^{2}}{1+z}\left(\frac{3 \operatorname{Li}_{3}(-z)}{2}+\operatorname{Li}_{3}(z)+\operatorname{Li}_{3}\left(\frac{1}{1+z}\right)-\frac{\operatorname{Li}_{2}(-z) \log (z)}{2}\right.\right.\notag\\
&- \frac{\operatorname{Li}_{2}(z) \log (z)}{2}-\frac{1}{24} \log ^{3}(z)-\frac{1}{6} \log ^{3}(1+z)+\frac{1}{4} \log (1+z) \log ^{2}(z) \notag\\
&+\left.\frac{\pi^{2}}{12} \log (1+z)-\frac{3 \zeta_{3}}{4}\right)+(1-z)\left(\frac{\operatorname{Li}_{2}(z)}{2}+\frac{1}{2} \log (1-z) \log (z)+\frac{15}{8}\right) \\
&-\left.\frac{1}{2}(1+z)\left(\operatorname{Li}_{2}(-z)+\log (z) \log (1+z)\right)+\frac{\pi^{2}}{24}(z-3)+\frac{1}{8}(11 z+3) \log (z)\right\} \,,\notag
\end{align}
\end{spreadlines}
\begin{spreadlines}{1.5ex}
\begin{align} 
C_{q g}^{(1,1)}(z)&= e_q^2 \Bigg\{\left(2 z^{2}-2 z+1\right)\left(\zeta_{3}-\frac{\operatorname{Li}_{3}(1-z)}{8}-\frac{\operatorname{Li}_{3}(z)}{8}+\frac{1}{8} \operatorname{Li}_{2}(1-z) \log (1-z)\right.\notag \\
&\left. +\frac{\operatorname{Li}_{2}(z) \log (z)}{8}-\frac{1}{48} \log ^{3}(1-z)+\frac{1}{16} \log (z) \log ^{2}(1-z)+\frac{1}{16} \log ^{2}(z) \log (1-z)\right) \notag\\
 & -\frac{3 z^{2}}{8}-\frac{1}{96}\left(4 z^{2}-2 z+1\right) \log ^{3}(z)+\frac{1}{64}\left(-8 z^{2}+12 z+1\right) \log ^{2}(z)\\ 
&+\frac{1}{32}\left(-8 z^{2}+23 z+8\right) \log (z)+\frac{5}{24} \pi^{2}(1-z) z+\frac{11 z}{32}+\frac{1}{8}(1-z) z \log ^{2}(1-z) \notag\\ 
&-\frac{1}{4}(1-z) z \log (1-z) \log (z)-\frac{1}{16}(3-4 z) z \log (1-z)-\frac{9}{32}
 \notag\\ 
&-\frac{1}{4}\left[z \log z+\frac{1}{2}\left(1-z^{2}\right)+\left(\pi^{2}-8\right) z(1-z)\right]
\Bigg\}\,,\notag
\end{align}
\end{spreadlines}
\begin{equation}
C_{q {\gamma}}^{(1,1)}(z) = {2 C_F C_A} C_{q {g}}^{(1,1)}(z) \,.
\end{equation}

The results above provide all the ingredients needed for the application of the $q_T$-subtraction formalism to the calculation of mixed \qcdqed corrections. The same coefficients are required by the transverse-momentum resummation formalism, considering in this case the full dependence on the resummation scale $Q$.

In the following section we present our phenomenological results for the case of $Z$ boson production.

\section{Phenomenological results}
\label{sec:results}

In order to obtain quantitative results, our calculation is implemented in two independent parton-level Monte Carlo programs. One of them is based on {\sc MCFM}~\cite{Boughezal:2016wmq} (including the NNLO QCD corrections), suitably modified to deal with mixed corrections and to apply the $q_T$-subtraction formalism. The other is a private implementation, which relies on the FKS subtraction method~\cite{Frixione:1995ms} to deal with the NLO-type divergencies (adapted to the mixed \qcdqed case), and on analytic results for the relevant scattering amplitudes obtained from ref.~\cite{Dixon:1998py}, plus an explicit calculation of the tree-level all-quarks channels using {\sc FeynCalc 9.2.0}~\cite{Shtabovenko:2016sxi}.

For our phenomenological analysis we consider $n_F=5$ massless quark flavours. We work in the $G_\mu$ scheme for the EW couplings, using the input values $G_\mu = 1.16639 \times 10^{-5}\text{~GeV}^{-2}$, $M_Z = 91.1876$~GeV and $M_W=80.385$~GeV. The width of the $Z$ boson is set to the value $\Gamma_Z = 2.4952$~GeV. For the parton luminosities and strong coupling, we use the NNPDF3.1luxQED set with five  flavours~\cite{Bertone:2017bme} through the LHAPDF interface~\cite{Buckley:2014ana}, always at NNLO accuracy, regardless the order of the calculation. Both  renormalisation and factorisation scales are set to the default value $\mu_R=\mu_F=m_{\ell_1\ell_2}$.
For the cutoff parameter of the subtraction method, $q_{T,\text{cut}}$, we choose the central value $q_{T,\text{cut}} = 0.2$~GeV. We checked that our results are compatible within uncertainties when varying this parameter around its central value by a factor of 2.

% version vieja:
%As a first check of our implementation, we compute the inclusive cross section for the production of an on-shell $Z$ boson, and compare to the results presented in ref.~\cite{deFlorian:2018wcj}, finding complete agreement. Splitting into quark-quark, quark-boson and boson-boson initiated channels, we find that the ${\cal O}(\asa)$ contributions to the total cross section are $(52.6 \pm 0.4)$~pb, $(-36.4 \pm 1.4)$~pb and $(0.58 \pm 0.01)$~pb, respectively, to be compared to the results $52.3$~pb, $-36.5$~pb and $0.57$~pb from the implementation in ref.~\cite{deFlorian:2018wcj}.
% version nueva con los numeros aca:
%As a first check of our implementation, we compute the inclusive cross section for the production of an on-shell $Z$ boson, and compare to the results presented in ref.~\cite{deFlorian:2018wcj}, finding complete agreement. Splitting into quark-quark, quark-gluon, quark-photon and gluon-photon initiated channels, we find that the ${\cal O}(\asa)$ contributions to the total cross section are $(52.6 \pm 0.4)$~pb, $(-34.8 \pm 0.3)$~pb, $(-1.41 \pm 0.01)$~pb and $(0.57 \pm 0.01)$~pb, respectively, to be compared to the results $52.3$~pb, $-35.0$~pb, $-1.41$~pb and $0.57$~pb from the implementation in ref.~\cite{deFlorian:2018wcj} (the numerical uncertainties obtained from the inclusive implementation are lower than the last digits shown).
% version nueva con los numeros en la tabla:
As a first check of our implementation, we computed the inclusive cross section for the production of an on-shell $Z$ boson, and compared to the predictions obtained from the analytic results presented in ref.~\cite{deFlorian:2018wcj}. The corresponding ${\cal O}(\asa)$ contributions to the cross section, split into quark-quark, quark-gluon, quark-photon and gluon-photon initiated channels, are shown in table~\ref{table:totalXS}. As can be seen from the table, we can reach sub-percent precision for these inclusive predictions, and we find full agreement with the analytic results from ref.~\cite{deFlorian:2018wcj}.
As an additional validation, we have computed the NNLO QCD differential distributions using the public code {\sc Matrix}~\cite{Grazzini:2017mhc}, finding full agreement with our results.

{\renewcommand{\arraystretch}{1.6}
\begin{table}[t]
\begin{center}
%\begin{tabular}{|l|c|c|c|c|c|}
%\begin{tabular}{l|C{2.5cm}|C{2.5cm}|C{2.5cm}|C{2.5cm}}
\begin{tabular}{l|S[table-format=2.1,table-column-width=2.5cm]|S[table-format=2.1,table-column-width=2.5cm]|S[table-format=1.2,table-column-width=2.5cm]|S[table-format=1.3,table-column-width=2.5cm]}
\hline
 Channel & $q q'$ & $qg$ & $q\gamma$ & $g\gamma$ \\ \hline
  $q_T$-subtraction [pb] & 52.6(4) & -34.8(3) & -1.41(1) & 0.569(2) \\
 Analytic (ref.~\cite{deFlorian:2018wcj}) [pb] & 52.3 & -35.0 & -1.41 & 0.571 \\
 \hline
\end{tabular}
\end{center}
\vspace*{-0.4cm}
\caption{
The ${\cal O}(\asa)$ contribution to the inclusive on-shell $Z$ production cross section for the different partonic channels. The results obtained using $q_T$-subtraction are compared to the inclusive predictions obtained in ref.~\cite{deFlorian:2018wcj}. Numerical uncertainties on the last digit are indicated in parenthesis for our predictions, while the uncertainties of the inclusive implementation are below the last digits shown. The category denoted by $qq'$ includes all combinations of quarks and anti-quarks.
}
\label{table:totalXS}
\end{table}
}

For all of the differential distributions presented here, we consider the following set of cuts,
\begin{equation}\label{eq:cuts}
p_{T,\ell_1}>25\, {\rm GeV} \, , \,\,\,\,  p_{T,\ell_2}>20\, {\rm GeV} 
\, , \,\,\,\,  |y|_{\ell_{1,2}}<2.5\, , \,\,\,\,  m_{\ell_1 \ell_2}>50\, {\rm GeV,}
\end{equation}
where $\ell_1$ and $\ell_2$ represent the final-state leptons, ordered according to their transverse momentum.
Since we consider only neutrinos in the final state, there is no need to recombine  collinear leptons and photons.

\begin{figure}[t]
     \centering
     \includegraphics[width=0.48\textwidth]{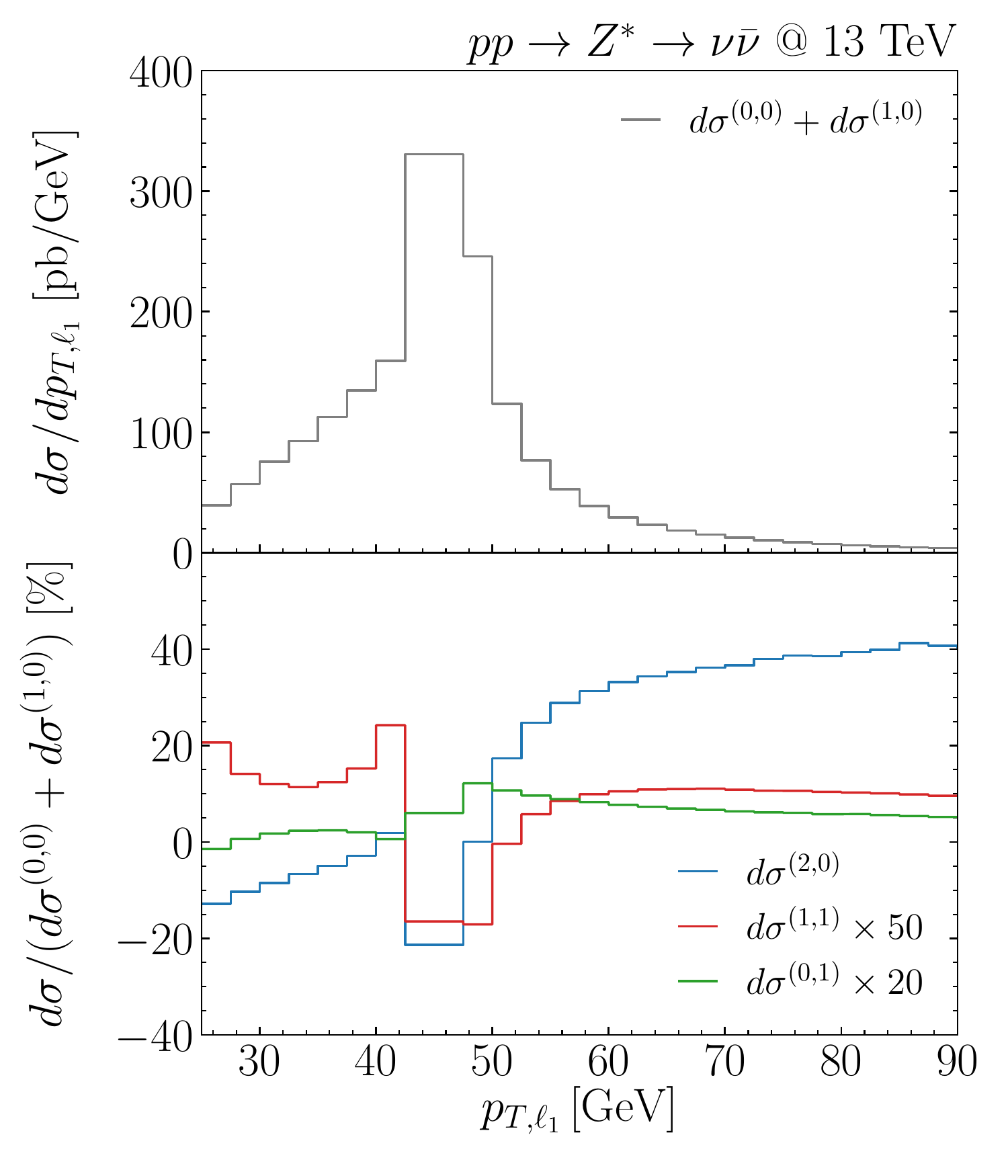}
     \includegraphics[width=0.48\textwidth]{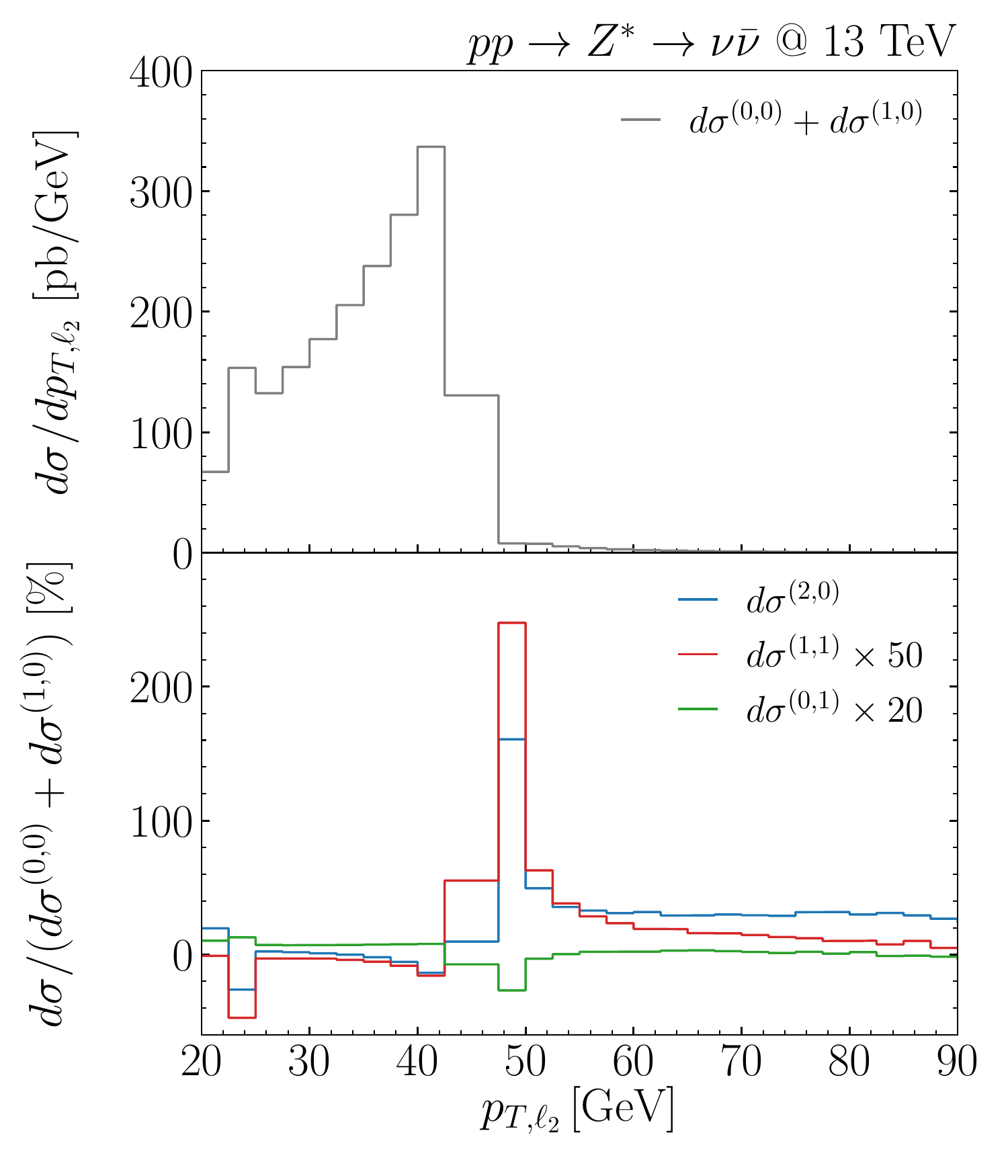}
     \vspace*{-0.4cm}
     \caption{Transverse momentum distributions for the hardest (left) and softer (right) lepton. The upper panel shows the NLO QCD prediction, while the lower panel shows the NNLO QCD (blue), NLO QED (green) and mixed (red) corrections, normalized to the NLO result.}
     \label{fig:ptlep}
\end{figure}

We start by presenting the transverse momentum distribution of the leptons in figure~\ref{fig:ptlep}.
The kinematical dependence of the mixed corrections is highly non trivial. This feature is also shared by the pure QCD and QED corrections, and it is expected due to the particular features that these two distributions present at fixed order in perturbation theory. At LO both leptons are back-to-back, and therefore the distributions are identical. The radiative corrections produce the change of shape that render the $p_{T,\ell_1}$ spectrum harder than the $p_{T,\ell_2}$ one, producing therefore sizeable distortions in the distribution. Furthermore, some regions of the phase space are almost not populated at LO, and therefore radiative corrections become more relevant. This is the case for the region of $p_{T,\ell_2}$ below the lower cut on $p_{T,\ell_1}$, which is directly not allowed for Born kinematics, or the region above $p_{T,\ell_{1,2}} \simeq M_Z/2$, which does not receive contributions from the $Z$ peak at LO.

\begin{figure}[t]
     \centering
     \includegraphics[width=0.48\textwidth]{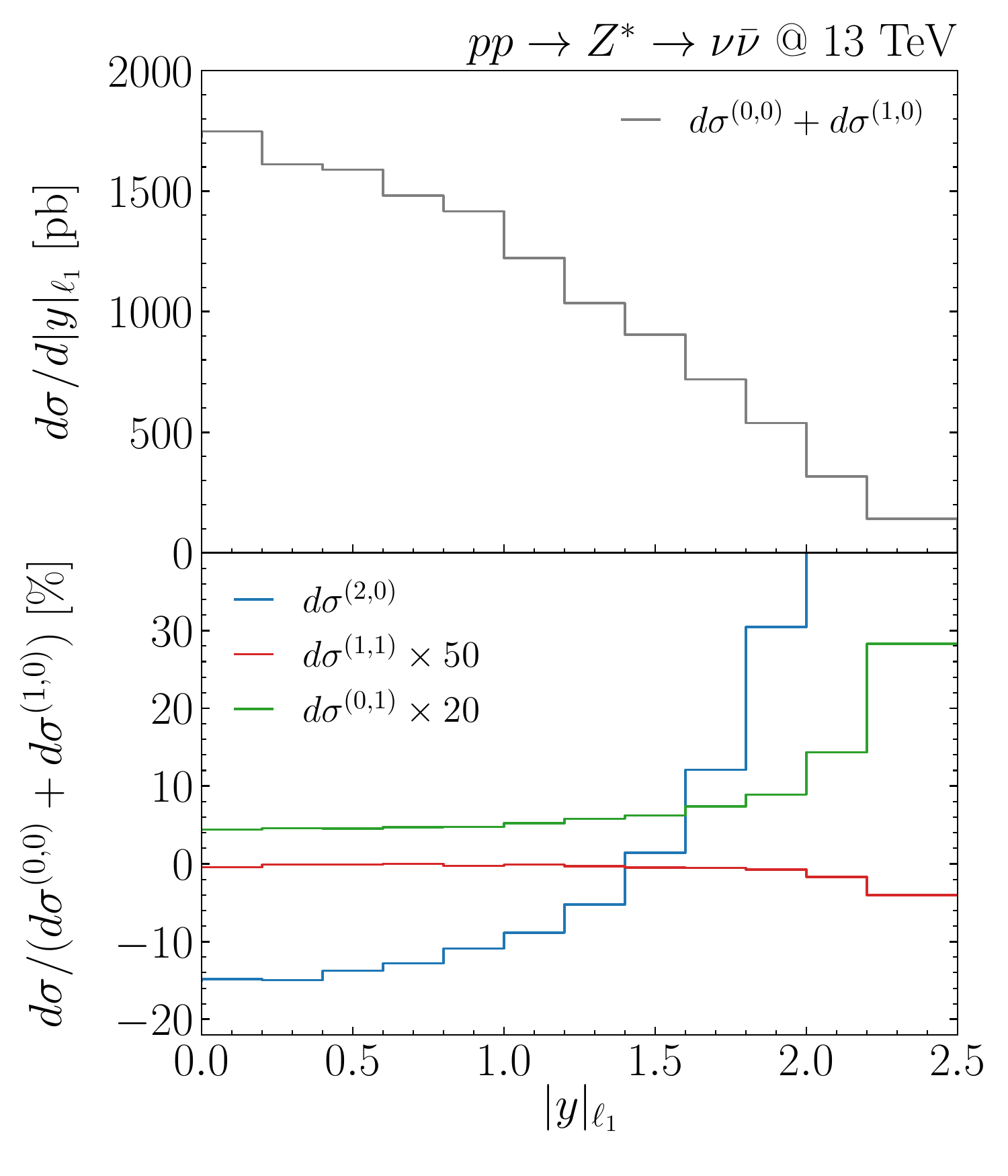}
     \includegraphics[width=0.48\textwidth]{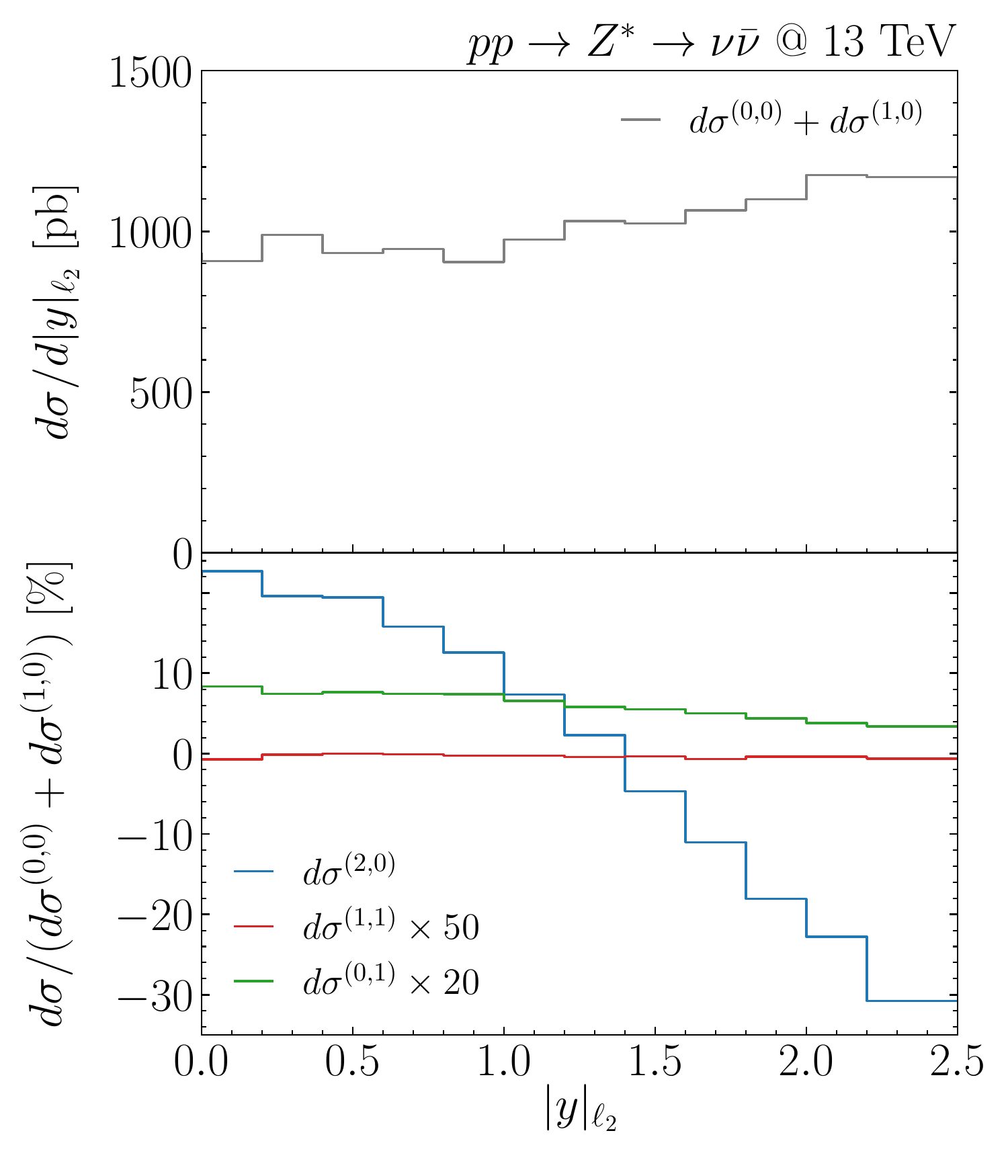}
     \vspace*{-0.4cm}
     \caption{Rapidity distributions for the hardest (left) and softer (right) lepton. The upper panel shows the NLO QCD prediction, while the lower panel shows the NNLO QCD (blue), NLO QED (green) and mixed (red) corrections, normalized to the NLO result.}
     \label{fig:raplep}
\end{figure}

From figure~\ref{fig:ptlep} we can observe that for $p_{T,\ell_1}<M_Z/2$ the mixed corrections are positive, representing an increase of about $0.5\%$ with respect to the NLO prediction. The corrections then change  sign, being of the order of $-0.5\%$ in the first bins after $p_{T,\ell_1}=M_Z/2$, which corresponds to the expected Sudakov shoulder near the kinematic boundaries mentioned in the previous paragraph \cite{Catani:1997xc}.
The mixed corrections then result smaller at the tail of the distribution.
With respect to the softer lepton, we can observe that the corrections become very large around and slightly above $p_{T,\ell_2}=M_Z/2$, a pattern  shared by the NNLO QCD corrections. In this region, the effect of the mixed \qcdqed contribution can reach the ${\cal O}(5\%)$ with respect to the NLO QCD result.
In addition, we can also observe a small (negative) peak in the corrections around $p_{T,\ell_2}=25$~GeV, which is related to the presence of a cut in $p_{T,\ell_1}$, as mentioned before.

We continue by presenting the rapidity distributions of the leptons, again ordered according to their transverse momentum, in figure~\ref{fig:raplep}.
In both cases, we can observe that the mixed corrections are extremely small, and show a very mild dependence on the corresponding kinematical variable. The reason for this particularly small value of the corrections is a very strong cancellation between the main partonic channels, that is the $q\bar q$ and $qg$ initiated processes, over the whole rapidity range under consideration, a pattern that can also be observed for instance at the level of the total cross section. 
We note that this effect is even stronger with the set of cuts in eq.~(\ref{eq:cuts}), compared to the fully inclusive case, with cancellations of about $90\%$ between the different channels.

\begin{figure}[t]
     \centering
     \includegraphics[width=0.48\textwidth]{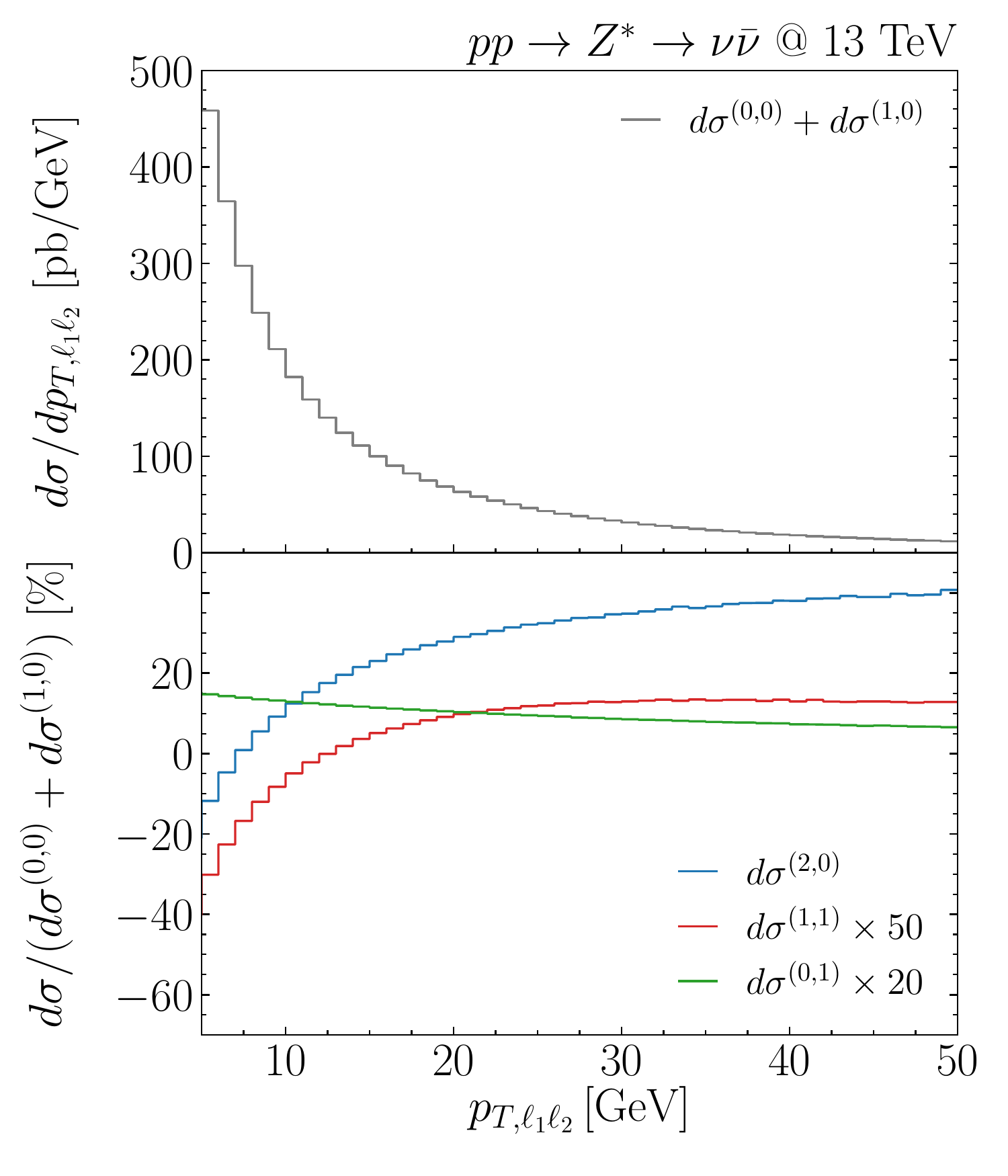}
     \includegraphics[width=0.48\textwidth]{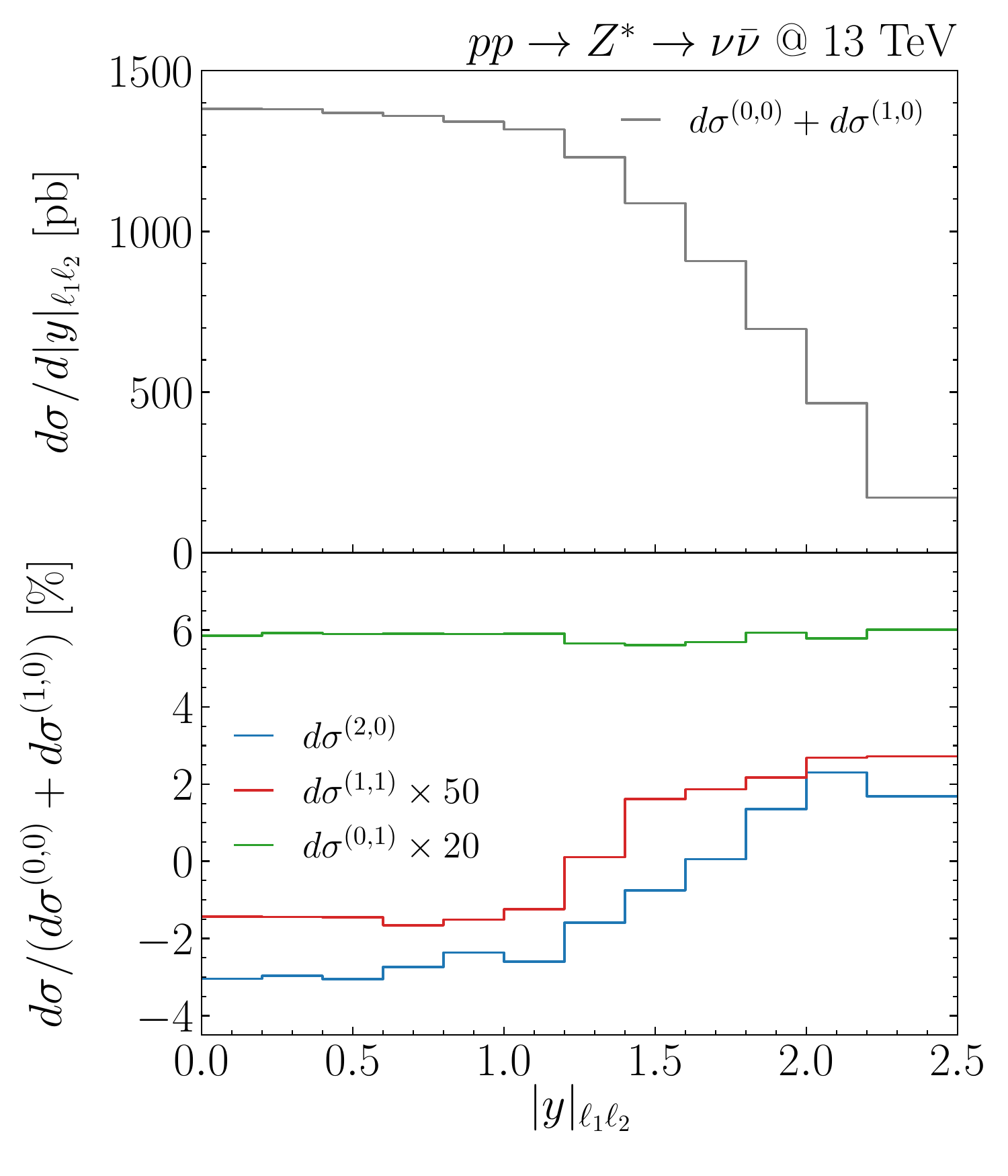}
     \vspace*{-0.4cm}
     \caption{Lepton-pair transverse momentum (left) and rapidity (right) distributions. The upper panel shows the NLO QCD prediction, while the lower panel shows the NNLO QCD (blue), NLO QED (green) and mixed (red) corrections, normalized to the NLO result.
     }
     \label{fig:ptz_and_yz}
\end{figure}
 
In figure~\ref{fig:ptz_and_yz} we present distributions for the lepton-pair system, specifically its transverse momentum and rapidity.
The mixed corrections are negative below $p_{T,\ell_1\ell_2} \sim 15$~GeV, and diverge in the $p_{T,\ell_1\ell_2} \to 0$ limit. 
The sign of the mixed corrections in the low transverse momentum region is the same as the one of the NNLO QCD corrections, as one can infer from the sign of the logarithmic coefficient with highest power (see eq.~(\ref{eq:Sig11_24}) for the mixed corrections and eq.~(66) of ref.~\cite{Bozzi:2005wk} for NNLO QCD).
Above $p_{T,\ell_1\ell_2} \sim 15$~GeV the mixed corrections become positive, increasing the NLO QCD result by about $0.3\%$. In the same region the NLO QED corrections are of the order of $0.5\%$. As it is well known, at low-$q_T$, the large logarithmic corrections to the cross section have to be treated with transverse momentum resummation in order to recover the reliability of the prediction. This is true not only for the transverse momentum distribution but for any observable which presents a kinematical region directly related to $q_T=0$.

The mixed corrections for the lepton-pair rapidity present a kinematic dependence that is similar to the one of the NNLO QCD contribution. They are negative for small $|y|_{\ell_1\ell_2}$, and become positive for larger values of rapidity. The overall size of the mixed corrections is of course much smaller though, being of the order of 50 times smaller than the NNLO QCD corrections.

\begin{figure}[t]
     \centering
     \includegraphics[width=0.48\textwidth]{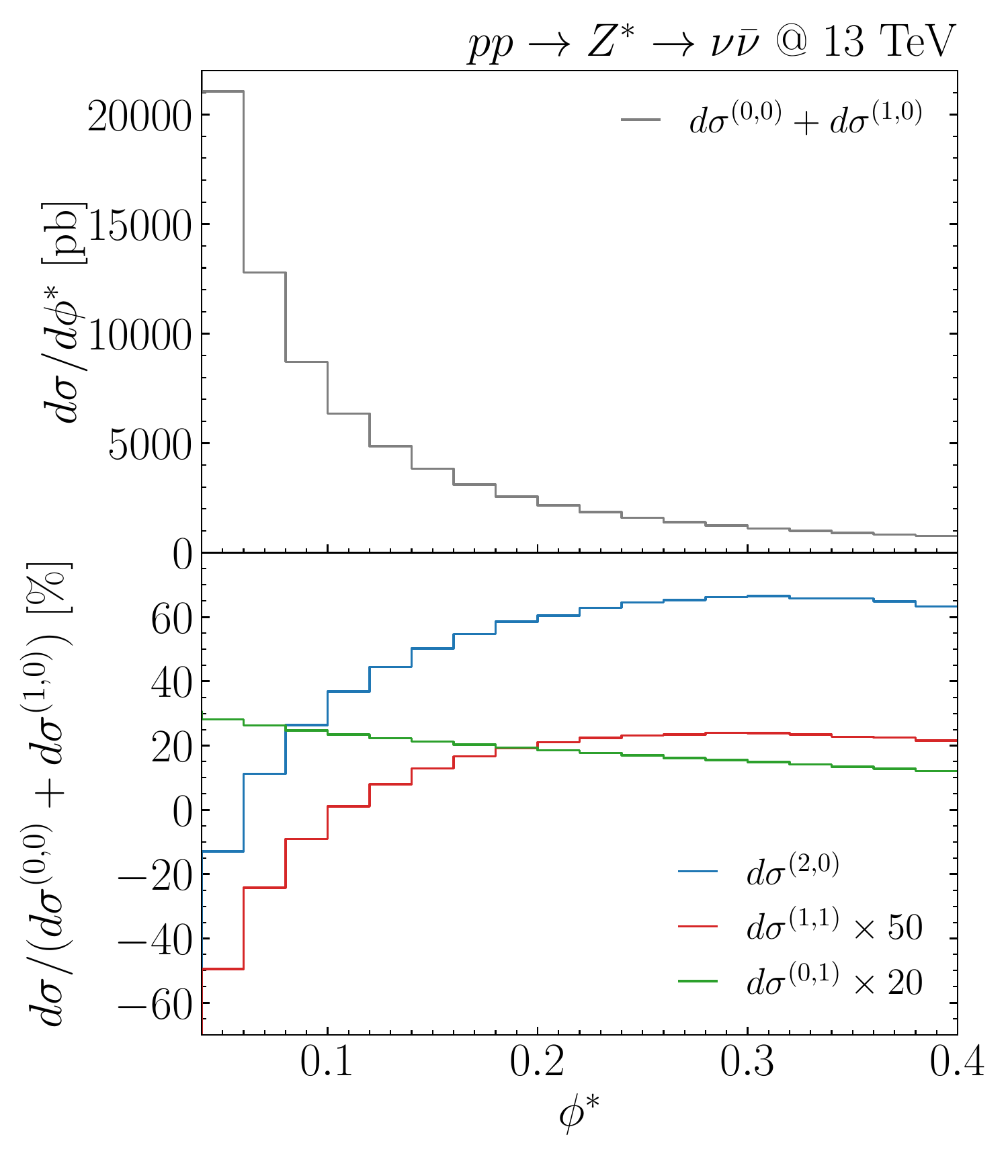}
     \includegraphics[width=0.48\textwidth]{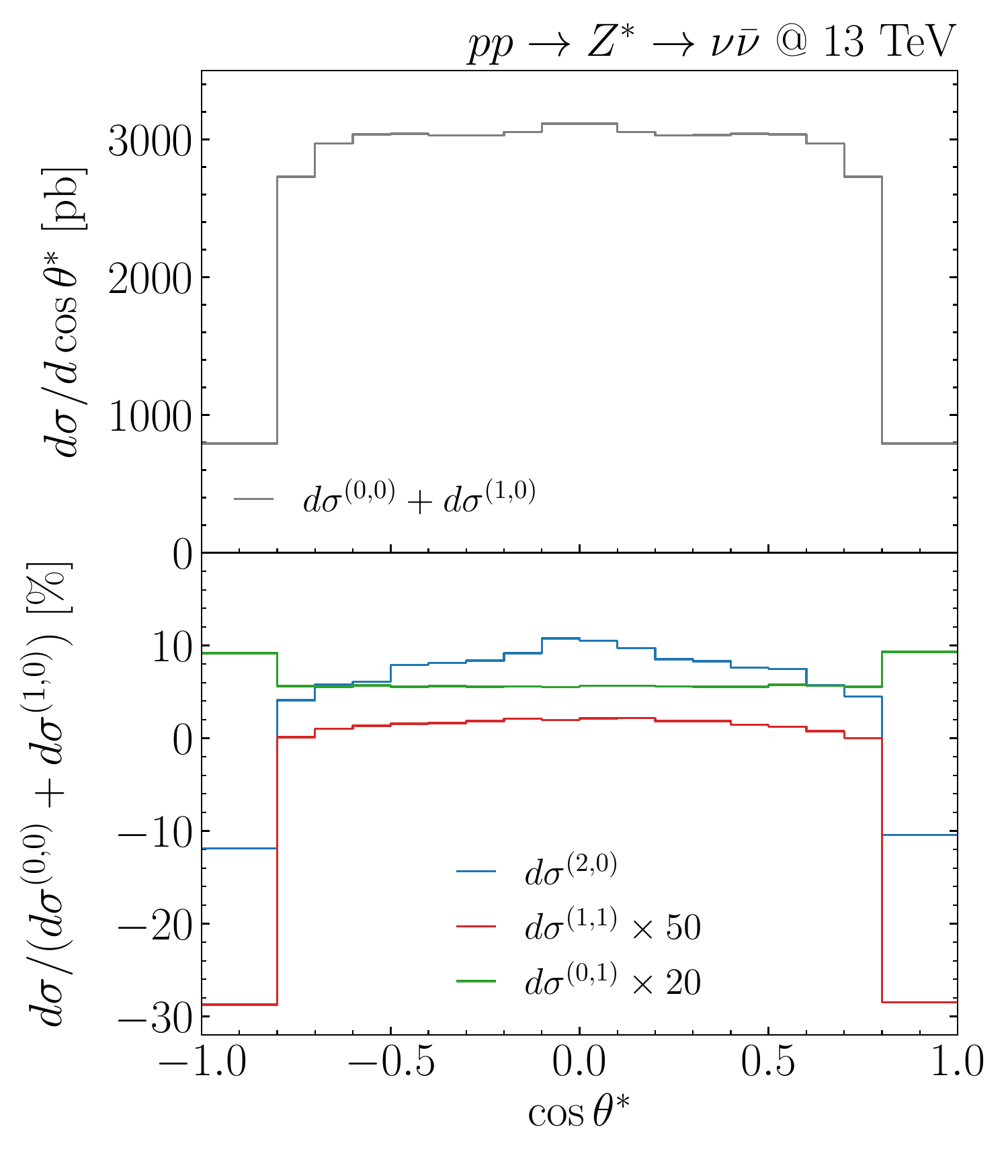}
     \vspace*{-0.4cm}
     \caption{The $\phi^*$ distribution. The upper panel shows the NLO QCD prediction, while the lower panel shows the NNLO QCD (blue), NLO QED (green) and mixed (red) corrections, normalized to the NLO result.
     }
     \label{fig:phi}
\end{figure}

Finally, we present in figure~\ref{fig:phi} the $\phi^*$ and $\cos\theta^*$ distributions, defined as \cite{Alioli:2016fum} 
\begin{align}
 &\phi^*= \tan\left(\frac{\pi - \Delta \Phi}{2}\right) \sin\theta^*  \nonumber\\
 &\Delta \Phi= \phi_{\,\ell_1}-\phi_{\,\ell_2}\\
 &\cos\theta^*= \tanh \left( \frac{y_{\,\ell_1}-y_{\,\ell_2}}{2}\right)\,.\nonumber
\end{align}

Since at LO the two leptons are back-to-back, the $\phi^*$ distribution is trivial at that order, and contributions with $\phi^* \neq 0$ only start at NLO. 
As in the case of the transverse momentum, the small-$\phi^*$ region is not well behaved at fixed order and it is necessary the use of transverse resummation in order to recover the reliability of the prediction in those kinematical regions.
The pattern of corrections, not only for the mixed but also for the NNLO QCD and NLO QED contributions, is very similar to the one observed in the $p_{T,\ell_1\ell_2}$ distribution, in particular with the mixed corrections being negative at small $\phi^*$ and becoming positive for larger values, and about a factor of 2 smaller than the NLO QED corrections in the tail of the distribution.

In the case of $\cos\theta^*$, the distribution is rather flat in the central region, and presents a strong suppression for $\cos\theta^* = \pm 1$, which is only populated by events with very large and opposite rapidities of the corresponding leptons. This region is therefore particularly suppressed by the presence of the cuts on $y_{\,\ell_{1,2}}$, which directly forbid the region above $|\cos\theta^*| \sim 0.987$. From the lower panel of the figure we can observe that the perturbative corrections are rather flat in the region where the bulk of the cross section is located, and therefore they follow a pattern similar to the one observed for the total cross section. In particular, the mixed \qcdqed corrections are extremely small, and become more relevant only close to the boundaries, where they reach the $0.6\%$ level (note that the last bin of the distribution is larger and extends from $|\cos\theta^*|=0.8$ to 1).

\begin{figure}[t]
     \centering
     \includegraphics[width=0.325\textwidth]{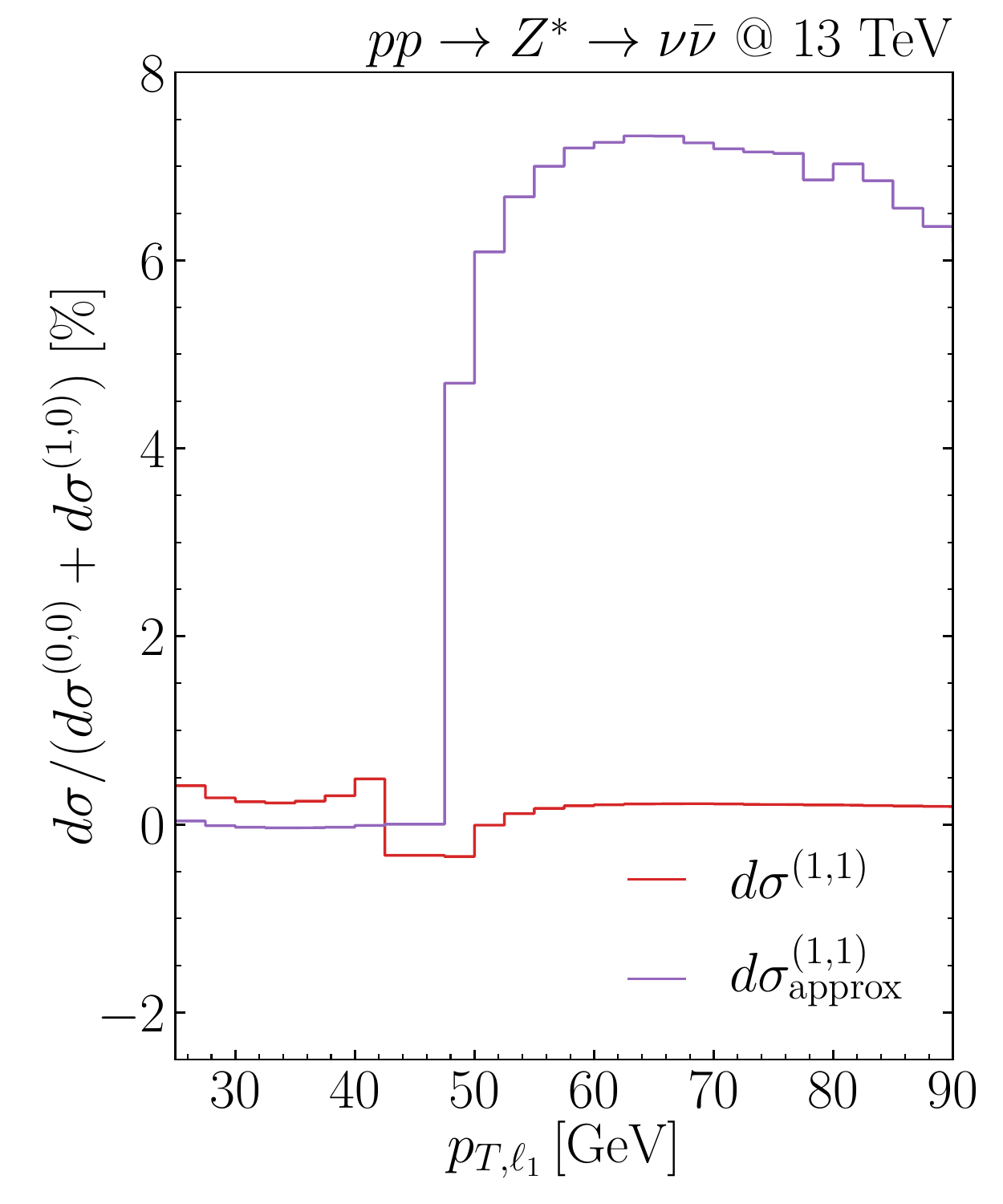}
     \includegraphics[width=0.325\textwidth]{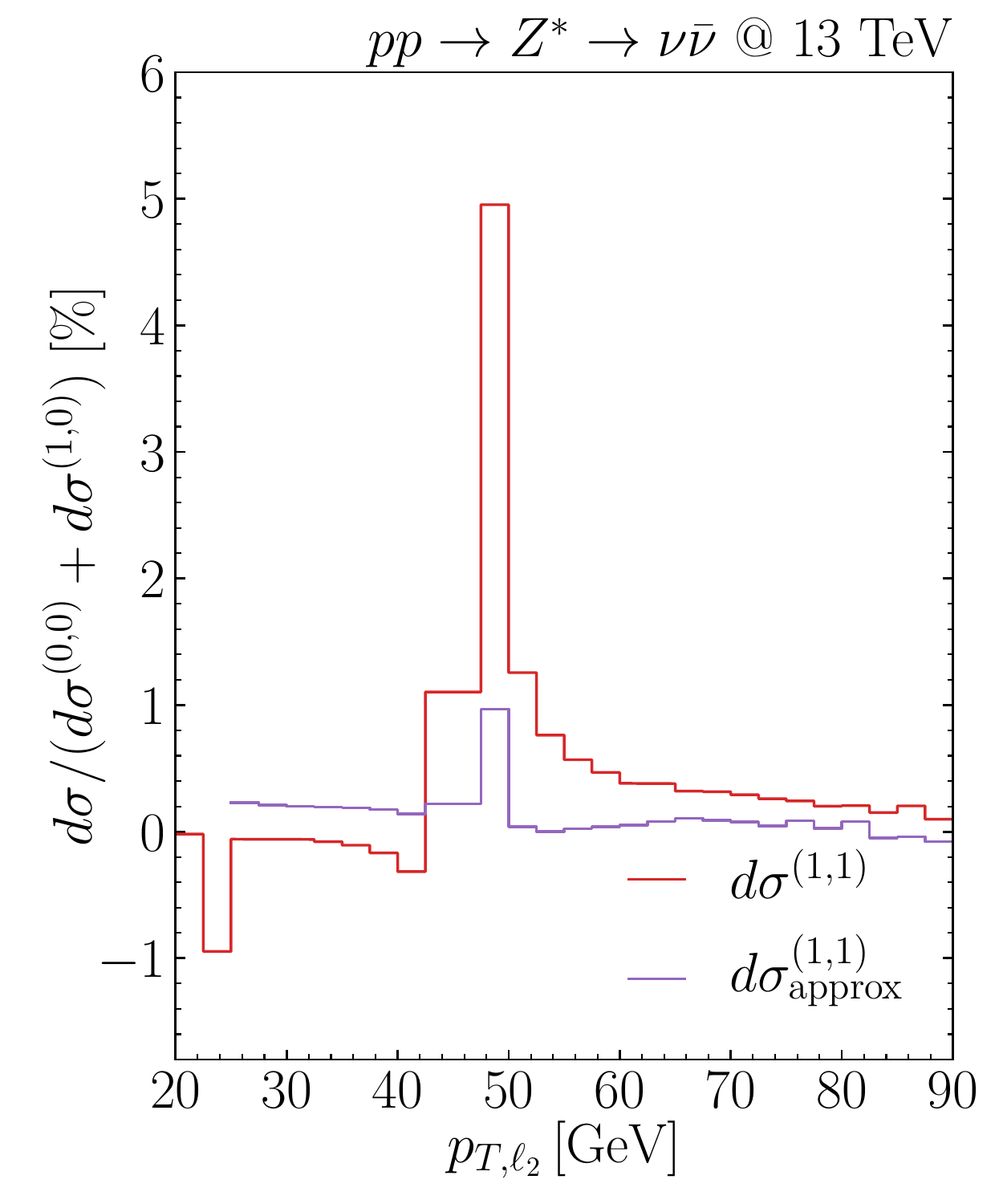}
     \includegraphics[width=0.325\textwidth]{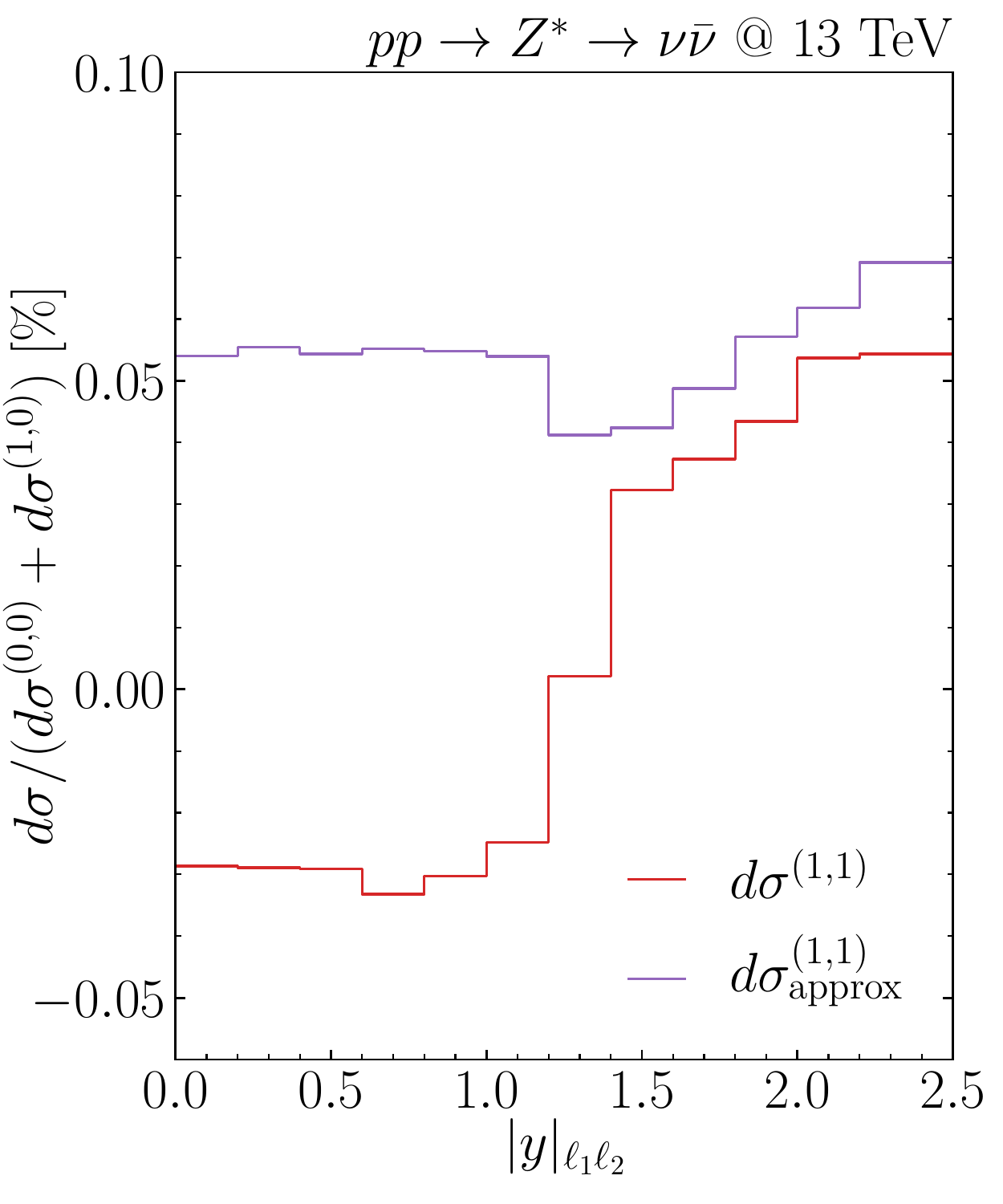}
     \vspace*{-0.3cm}
     \caption{Comparison between the mixed \qcdqed corrections (red) and the naive factorisation approximation (purple), for the transverse momentum of the hardest (left) and softer (center) lepton, and the rapidity of the pair (right).
     }
     \label{fig:comparison}
\end{figure}

Before going to the summary, it is interesting to compare the size of the mixed \qcdqed corrections computed here against the naive approximation in which QCD and QED corrections factorize. Specifically, defining for a given bin
\begin{equation}
    d\Delta^{(i,j)} = d\sigma^{(i,j)}/d\sigma^{(0,0)}\,,
\end{equation}
the multiplicative approximation to the ${\cal O}(\asa)$ based on NLO QCD and QED predictions is given by the product
\begin{equation}\label{eq:approx}
    d\sigma^{(1,1)}_\text{approx} = d\sigma^{(0,0)} d\Delta^{(1,0)} d\Delta^{(0,1)}\,.
\end{equation}

In figure~\ref{fig:comparison} we present the mixed
\qcdqed corrections together with the approximation defined by eq.~(\ref{eq:approx}), for the transverse momentum of the two leptons and the rapidity of the pair. The results are normalized to the NLO QCD prediction, as in the lower panels of the previous figures.
We can observe that, in all cases, the multiplicative approach is a rather poor approximation to the full results. This is in line with the observations made for the total cross section in ref.~\cite{deFlorian:2018wcj}.
The discrepancies, however, can be strongly enhanced at the differential level. This can be seen for instance in the $p_{T,\ell_1} > M_Z/2$ region, where the exact ${\cal O}(\asa)$ corrections are at the per-mille level, while the factorisation approximation predicts $\sim7\%$ corrections.
The reason for this big discrepancy is the presence of large $K$-factors at NLO (both in QCD and QED), associated to the fact that at LO this region is only populated by events that are away from the $Z$ peak.
In the case of the $p_{T,\ell_2}$ distribution, we can observe that the multiplicative approach has the wrong sign for $p_{T,\ell_2}<M_Z/2$ (note that the approximation is not well defined for $p_{T,\ell_2}<25$~GeV due to the cut in the hardest lepton), and fails to reproduce the correct size of the corrections around the peak located in $p_{T,\ell_2}\sim M_Z/2$. Finally, for the rapidity of the lepton pair we can see that the factorisation approximation predicts a rather flat $K$-factor, failing to describe the kinematical dependence of the mixed corrections.

\section{Summary}
\label{sec:conc}
By using the abelianisation techniques~\cite{deFlorian:2018wcj,deFlorian:2015ujt,deFlorian:2016gvk}, in this work we have extended the $q_T$-subtraction formalism in order to deal with the case of mixed \qcdqed corrections. The method can be applied to the fully exclusive calculation of the ${\cal O}(\asa)$ corrections for the production of a colourless and neutral final state (e.g. $Z$ and Higgs bosons, photons, neutrinos).
We have provided all the relevant formulas for its implementation at ${\cal O}(\asa)$. The coefficient functions and the hard virtual coefficients are also of value for transverse momentum resummation and our expressions contain the full dependence on the resummation scale $Q$. 

We have applied the method to the production of an off-shell $Z$ boson, and considered its decay into a pair of neutrinos. We presented differential distributions for the final-state leptons at the LHC, and found that the corrections can have a sizeable dependence on the kinematics, and not necessarily following the pattern of the NNLO QCD corrections for instance. The size of the corrections is typically very small and below $1\%$, though it can be enhanced in some particular phase space regions.
We note that our predictions are in qualitative agreement with the corresponding results in ref.~\cite{Delto:2019ewv}.

We have also compared the mixed \qcdqed contribution with the factorisation approximation based on the product of QCD and QED $K$-factors. We have found that this multiplicative approach is in general a bad approximation to the mixed corrections, and the disagreement can be quite extreme for some differential distributions.

As a final remark, it is interesting to point out that recent developments have allowed the application of the $q_T$-subtraction method to the production of a heavy-quark pair at NNLO in QCD~\cite{Bonciani:2015sha,Catani:2019iny,Catani:2019hip} (see also its related application to NLO EW corrections for massive lepton pair production in ref.~\cite{Buonocore:2019puv}). Following similar abelianisation techniques to the ones used in the present paper, the method could be extended to also deal with the mixed \qcdqed corrections for the production of a massive charged (colourless) final state.

\section*{Acknowledgements}
We thank Stefano Catani and Massimiliano Grazzini for discussions. LC would like to thank German Sborlini and Giancarlo Ferrera for useful comments and discussions. This project has received funding from the European Union's Horizon 2020 research and innovation programme under the Marie Skłodowska-Curie grant agreement No 754496 and  Conicet and ANPCyT.

\bibliography{biblio}

\end{document}